\documentclass[twocolumn,A4]{article}
\usepackage[dvips]{graphics}
\begin{document}
\onecolumn
\begin{center}
{\bf{\Large Persistent current and low-field magnetic susceptibility 
in one-dimensional mesoscopic rings: Effect of long-range hopping}}\\
~\\
Santanu K. Maiti$^{1,2,*}$ \\
~\\
{\em $^1$Theoretical Condensed Matter Physics Division,
Saha Institute of Nuclear Physics, \\ 
1/AF, Bidhannagar, Kolkata-700 064, India \\
$^2$Department of Physics, Narasinha Dutt College,
129, Belilious Road, Howrah-711 101, India} \\
~\\
{\bf Abstract}
\end{center}
Persistent current and low-field magnetic susceptibility in single-channel
normal metal rings threaded by a magnetic flux $\phi$ are investigated within 
the tight-binding framework considering long-range hopping of electrons in the 
{\em shortest} path. The higher order hopping integrals try to reduce the
effect of disorder by delocalizing the energy eigenstates, and accordingly, 
current amplitude in disordered rings becomes comparable to that of an 
ordered ring. Our study of low-field magnetic susceptibility predicts that 
the sign of persistent currents can be mentioned precisely in mesoscopic 
rings with fixed number of electrons, even in the presence of impurity in 
the rings. For perfect rings, low-field current shows only the diamagnetic 
sign irrespective of the total number of electrons $N_e$. On the other hand, 
in disordered rings it exhibits the diamagnetic and paramagnetic natures 
for the rings with odd and even $N_e$ respectively.  
\vskip 1cm
\begin{flushleft}
\end{flushleft}
\vskip 4in
\noindent
{\bf ~$^*$Corresponding Author}: Santanu K. Maiti

Electronic mail: santanu.maiti@saha.ac.in 
\newpage
\twocolumn

\section{Introduction}

The phenomenon of persistent current in mesoscopic normal metal rings has 
generated a lot of excitement as well as controversy over the past years. 
In a pioneering work, B\"{u}ttiker, Imry and Landauer~\cite{butt} predicted 
that, even in the presence of disorder, an isolated one-dimensional metallic 
ring threaded by a magnetic flux $\phi$ can support an equilibrium persistent 
current with periodicity $\phi_0=ch/e$, the elementary flux quantum. Later, 
the existence of persistent current was further confirmed by several 
experiments~\cite{chand,maily,levy,deb,reul,jari,rab}. However, these 
experiments yield many results those are not well-understood theoretically 
even today~\cite{cheu1,cheu2,mont,bouc,alts,von,schm,ambe,abra,bouz,giam,
wain,avishai,weiden,orella1,kulik,mos}. The results of the single loop 
experiments are significantly different from those for the ensemble of 
isolated loops. Persistent currents with expected $\phi_0$ periodicity 
have been observed in isolated single Au rings~\cite{chand} and in a 
GaAs-AlGaAs ring~\cite{maily}. Levy {\em et al.}~\cite{levy} found 
oscillations with period $\phi_0/2$ rather than $\phi_0$ in an ensemble 
of $10^7$ independent Cu rings. Similar $\phi_0/2$ oscillations were 
also reported for an ensemble of disconnected $10^5$ Ag rings~\cite{deb} 
as well as for an array of $10^5$ isolated GaAs-AlGaAs rings~\cite{reul}. 
In an experiment, Jariwala {\em et al.}~\cite{jari} obtained both 
$\phi_0$ and $\phi_0/2$ periodic persistent currents for an array of thirty 
diffusive mesoscopic Au rings. Except for the case of nearly ballistic 
GaAs-AlGaAs ring~\cite{maily}, all the measured currents are in general 
one or two orders of magnitude larger than those expected from theory. 
The diamagnetic response of the measured $\phi_0/2$ oscillations of 
ensemble-averaged persistent currents near zero magnetic field also 
contrasts with most predictions~\cite{schm,ambe}. 

Free electron theory predicts that, at zero temperature, an ordered 
one-dimensional metallic ring threaded by a magnetic flux $\phi$ supports 
persistent current with maximum amplitude $I_0=ev_F/L$, where $v_F$ is 
the Fermi velocity of an electron and $L$ is the circumference of the 
ring. Metals are intrinsically disordered which tends to decrease persistent 
current, and calculations show that the magnitude of the currents reduces 
to $I_0l/L$, where $l$ is the elastic mean free path of the electrons. 
This expression remains valid even if one takes into account the finite 
width of the ring by adding contributions from the transverse channels, 
since disorder leads to a compensation between the channels~\cite{cheu2,mont}. 
However, measurements on single isolated mesoscopic rings~\cite{chand,maily} 
detected $\phi_0$-periodic persistent currents with amplitudes of the order 
of $I_0\sim ev_F/L$, (close to the value for an ordered ring). Though theory 
seems to agree with experiment~\cite{maily} only when disorder is weak, the 
amplitudes of the currents in single-isolated-diffusive gold 
rings~\cite{chand} are two orders of magnitude larger than the theoretical 
estimates. This discrepancy initiated intense theoretical activity, and it 
is generally believed that the electron-electron correlation plays an 
important role in the disordered rings~\cite{abra,bouz,giam}, though the 
physical origin behind this enhancement of persistent current is still 
unclear. 

In this article we investigate a detailed study of persistent current and 
low-field magnetic susceptibility in single channel rings in the 
tight-binding framework considering long-range hopping of electrons in the
{\em shortest} path. Our calculations show that in a disordered ring 
with higher order hopping integrals, current amplitude is comparable to 
that of an ordered ring. This is due to the fact that higher order hopping
integrals try to delocalize the energy eigenstates and thus compensate
the effect of disorder. In the rest part of this article, we describe the
dependences of the sign of low-field currents as a function of the total
number of electrons $N_e$, and also discuss the effect of temperature 
on these low-field currents.

The plan of the paper is as follow. Section $2$ relates the behavior of
persistent current in the presence of long-range hopping integrals and
clearly describes how current amplitude in disordered rings becomes 
comparable to that of an ordered ring. In Section $3$, we investigate 
the behavior of low-field magnetic susceptibility at absolute zero 
temperature both for ordered and disordered rings as a function of $N_e$.
Section $4$ focuses the effect of temperature on the low-field currents 
and determines the critical value of magnetic flux $\phi_c(T)$ where 
current changes its sign from the paramagnetic to diamagnetic phase for 
the rings with even number of electrons. Finally, the conclusions of our 
study can be found in Section $5$.

\section{Persistent current}

\begin{figure}[ht]
{\centering \resizebox*{5cm}{4cm}{\includegraphics{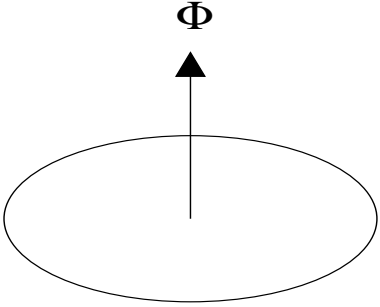}}\par}
\caption{One-dimensional mesoscopic normal metal ring threaded by a 
magnetic flux $\phi$. A persistent current $I$ is established in the ring.}
\label{ring}
\end{figure}

We describe a $N$-site ring (see Fig.~\ref{ring}) enclosing a magnetic flux 
$\phi$ (in units of the elementary flux quantum $\phi_0=ch/e$) by the 
following tight-binding Hamiltonian in the Wannier basis,
\begin{equation}
H=\sum_i\epsilon_i c_{i}^{\dagger}c_{i} +\sum_{i\ne j} v_{ij} 
\left[e^{-i \theta} c_{i}^{\dagger} c_{j}+ h.c. \right]
\label{hamil}
\end{equation} 
where $\epsilon_i$'s are the site energies, $v_{ij}$'s are the hopping 
integrals between the sites $i$ and $j$, and $\theta={2\pi\phi}(|i-j|)/N$. 
The long-range hopping (LRH) integrals are taken as, 
\begin{equation}
v_{ij}=\frac{v}{\left|\frac{N}{\pi} \sin\left[\frac{\pi}{N}(i-j)\right]
\right|^{\alpha}} 
\end{equation}
where, $v$ being a constant representing the nearest-neighbor hopping (NNH) 
integral. In the present work, electron-electron interaction is not included, 
and therefore, we do not consider the spin of the electrons since it will 
not change any qualitative behavior of persistent currents. Throughout 
our study we set $v=-1$, and use the units where $c=e=h=1$.

An electron in an eigenstate with energy $E_n$ carries a persistent
current $I_n(\phi)=-\partial E_n/\partial \phi$, and at zero
temperature total persistent current is given by $I(\phi)=\sum_n I_n(\phi)$,
where the summation is over all the states below the Fermi level. 

\begin{figure}[ht]
{\centering \resizebox*{7.5cm}{10cm}{\includegraphics{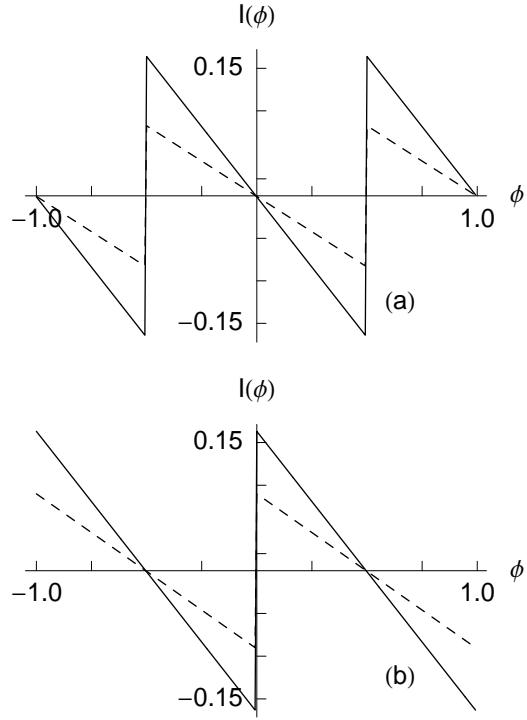}}\par}
\caption{Current-flux characteristics for some perfect rings with size 
$N=120$. The solid and dotted curves are respectively for the rings with LRH 
($\alpha=1.3$) and NNH integrals, where (a) $N_e=35$ (odd) and (b) $N_e=40$
(even).}
\label{longorder}
\end{figure}

For an ordered ring, setting $\epsilon_i=0$ for all $i$, the energy of the
$n$-th eigenstate can be expressed as,
\begin{equation}
E_n(\phi)=\sum_m \frac{2v}{m^{\alpha}}\cos\left[\frac{2\pi m}{N}
(n+\phi)\right]
\label{engy}
\end{equation}
where $m$ is an integer and it runs from $1$ to $N/2$ for the rings with 
even $N$, while it goes from $1$ to $(N-1)/2$ for those rings described 
by odd $N$. Now the persistent current carried by this $n$-th
eigenstate becomes,
\begin{equation}
I_n(\phi)=\left(\frac{4\pi v}{N}\right)\sum_m m^{1-\alpha}\sin\left[
\frac{2\pi m}{N}(n+\phi)\right]
\label{curr}
\end{equation}
For very large value of $\alpha$, Eqs.~(\ref{engy}) and (\ref{curr})
essentially reduce to the expressions for the energy spectrum and persistent
current of an ordered ring described by the nearest-neighbor tight-binding
Hamiltonian. As we decrease $\alpha$, contributions from the higher order
hopping integrals become appreciable which modify the energy spectrum and
persistent current, and we will see that the modifications are quite 
significant in the presence of disorder. Figure~\ref{longorder} shows the
variation of persistent current as a function of magnetic flux $\phi$ for 
some perfect rings ($N=120$). The solid and dotted curves represent the 
results for the rings described by LRH ($\alpha=1.3$) and NNH integrals 
respectively, where the curves plotted in (a) give the variation of the 
currents for the rings with odd $N_e$ 
($N_e=35$) and the same are plotted for the rings with even $N_e$ ($N_e$=40) 
in (b). The results predict that the current amplitude increases in the 
presence of LRH integrals, compared to the NNH integral. In this context 
it has been examined that, in the presence of LRH integrals, the amplitude 
of the current initially increases (not shown here in the figure) as we 
increase the ring 
size, but eventually it falls when the ring becomes larger. This is due
to the fact that as we increase the number of sites, the Hamiltonian
Eq.~(\ref{hamil}) includes additional higher order hopping integrals which 
cause an increase in the net velocity of the electrons, but after certain
ring size the increment in velocity drops to zero since the additional
hopping integrals are then between far enough sites giving negligible
contributions. 

Now we address the problem of persistent current in the presence of disorder.
In order to introduce the disorder in the ring, we choose the site energies
$\epsilon_i$'s randomly from a ``Box" distribution function of width $W$,
which reveal that the ring is subjected to the diagonal disorder. As 
representative examples of persistent current in disordered rings, we plot 
\begin{figure}[ht]
{\centering \resizebox*{7.5cm}{10cm}{\includegraphics{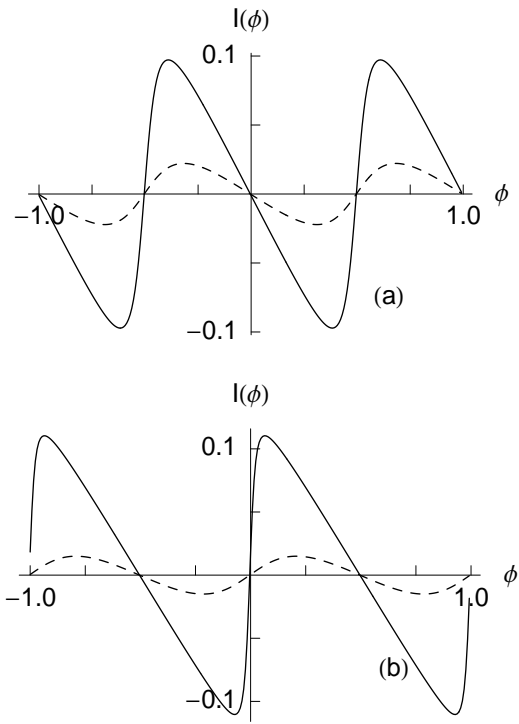}}\par}
\caption{Current-flux characteristics for some typical disordered rings 
($W=1$) with size $N=120$. The solid and dotted curves correspond to the 
rings with LRH ($\alpha=1.3$) and NNH integrals respectively, where 
(a) $N_e=35$ (odd) and (b) $N_e=40$ (even).}
\label{longdisorder}
\end{figure}
the results in Fig.~\ref{longdisorder} for some $120$-site rings taking 
$W=1$. All the curves shown in Fig.~\ref{longdisorder} are performed for 
the distinct disordered configurations of the rings and no averaging is done 
here since in the averaging process several mesoscopic phenomena disappear. 
The solid and dotted curves correspond to the rings with LRH and NNH 
integrals respectively, and our results predict that current amplitude gets 
an order of magnitude enhancement in the rings described by the LRH 
integrals compared to those rings described by the NNH integral. 
Figure~\ref{longdisorder}(a) shows the variation of the persistent current 
for the rings with odd $N_e$ ($N_e=35$) and Fig.~\ref{longdisorder}(b) gives 
the variation of the currents for the rings with even $N_e$ ($N_e=40$).
It is apparent from Figs.~\ref{longorder} and \ref{longdisorder} that the 
current amplitudes in disordered rings with LRH integrals are of the same 
order of magnitude as observed in ordered rings. We have also seen that 
the decrease in amplitude of the current is quite small even if we increase 
the strength of disorder. In the NNH models current amplitudes are 
suppressed due to the localization of the energy eigenstates~\cite{lee1}. 
On the other hand, the present tight-binding model with LRH integrals 
supports extended electronic eigenstates even in the presence of disorder 
and for this reason persistent currents are not reduced by the impurities.

\section{Magnetic susceptibility at $T=0$ K}

The sign of persistent currents can be determined exactly by calculating
magnetic susceptibility, and here we investigate the properties of low-field
\begin{figure}[ht]
{\centering \resizebox*{7.5cm}{5.5cm}{\includegraphics{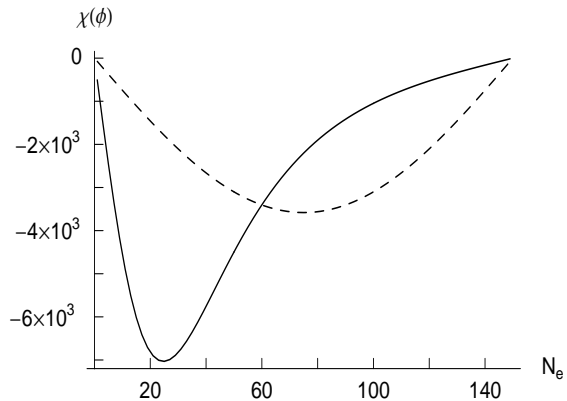}}
\par}
\caption{$\chi(\phi)$ versus $N_e$ curves for some perfect rings with 
$N=150$. The solid and dotted curves represent the rings with LRH 
($\alpha=1.4$) and NNH integrals respectively.}
\label{longsusceporder}
\end{figure}
currents for the rings with fixed number of electrons $N_e$. The general 
expression of magnetic susceptibility is expressed in the form,
\begin{equation}
\chi(\phi)=\frac{N^3}{16\pi^2}\left(\frac{\partial I(\phi)}{\partial \phi}
\right)
\end{equation}
Thus by calculating the sign of $\chi(\phi)$ we can predict whether the 
current is diamagnetic or paramagnetic.

Let us first discuss the properties of low-field currents in perfect rings
at absolute zero temperature ($T=0$ K). In Fig.~\ref{longsusceporder}, we 
plot $\chi(\phi)$ as a function of $N_e$ in the limit $\phi\rightarrow 0$ 
for some perfect rings with $N=150$. The solid and dotted curves represent 
the variation of $\chi$ for the rings described by LRH and NNH integrals
respectively. The results show that, for perfect rings, low-field currents 
exhibit only the diamagnetic sign irrespective of the total number of 
electrons $N_e$ in the ring. This variation can be clearly understood 
if we consider the slope of the curves as plotted in 
Figs.~\ref{longorder}(a) and (b) 
\begin{figure}[ht]
{\centering \resizebox*{7.5cm}{10cm}
{\includegraphics{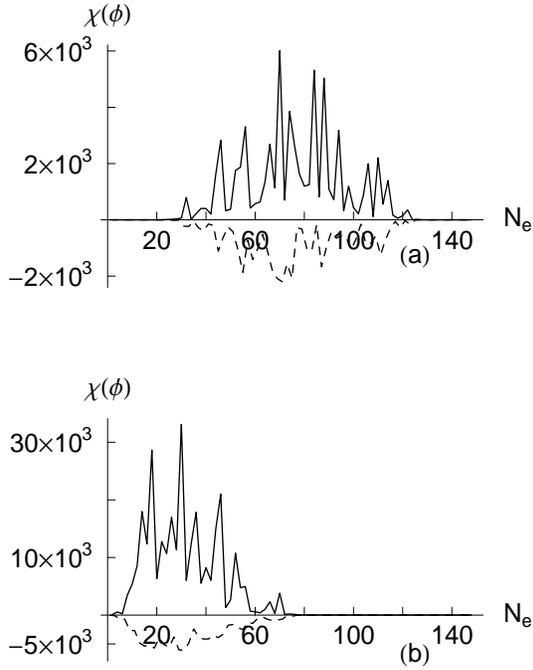}}\par}
\caption{$\chi(\phi)$ versus $N_e$ curves of some disordered rings ($W=1$) 
with size $N=150$. The solid and dotted lines are respectively for the 
rings with even and odd $N_e$, where (a) NNH and (b) LRH ($\alpha=1.4$) 
integrals.} 
\label{longsuscepdisorder}
\end{figure}
near $\phi=0$. Thus, both for the perfect rings with odd and even $N_e$, 
current has only negative slope which predicts the diamagnetic persistent 
current.

The effect of impurities on the sign of low-field currents is quite 
interesting. As representative examples, in Fig.~\ref{longsuscepdisorder}
we display the variation of $\chi$ as a function of $N_e$, where
(a) and (b) represent the rings with NNH and LNH integrals respectively. The 
solid and dotted lines correspond to the results for the rings containing 
even and odd number of electrons respectively. These results emphasize that, 
in the disordered rings the low-field currents exhibit the diamagnetic sign 
for odd $N_e$, while we get the paramagnetic response for the rings with 
even $N_e$. The diamagnetic and the paramagnetic natures of the low-field 
currents in the presence of impurity in the rings can be understood easily 
if we take the slope of the curves as given in Figs.~\ref{longdisorder}(a) and
(b). Such an effect of disorder on the low-field currents is true for any 
disordered configuration. Accordingly, in the presence of impurity one can 
easily predict the sign of the low-field currents both for the rings with 
odd and even $N_e$, irrespective of the specific realization of disordered 
configuration of the rings. 

\section{Magnetic susceptibility at finite temperature}

This section focuses the effect of temperature on the low-field currents. 
\begin{figure}[ht]
{\centering \resizebox*{7.5cm}{10cm}{\includegraphics{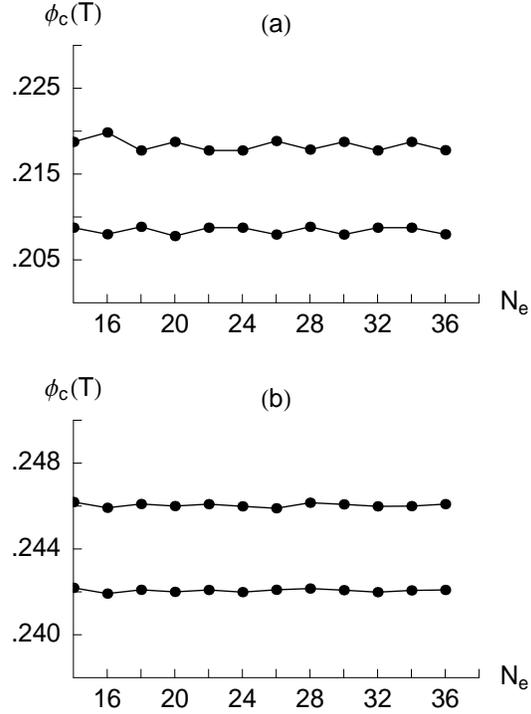}}\par}
\caption{Variation of $\phi_c(T)$ with $N_e$ (even $N_e$ only) for some
disordered rings ($W=1$) taking $N=60$, where (a) rings with NNH integrals 
and (b) rings with LRH ($\alpha=1.6$) integrals. The upper and lower curves 
both in (a) and (b) are respectively for the rings with $T/T^{\star}=1.0$ 
and $T/T^{\star}=0.5$.}
\label{longtemp}
\end{figure}
As temperature increases the probability that electrons occupy higher 
energy levels, those may carry larger currents, increases. But if we increase 
the temperature in such a way that crosses the energy gap between two 
successive energy levels, which carry currents in opposite directions,
then mutual cancellations of the positive and negative currents decrease the
net current amplitude. Therefore, it is necessary to specify a characteristic 
temperature $T^{\star}$, which is determined by the energy level spacing 
$\Delta$. At finite temperature, thermal excitations, such as phonons, 
are present which interact with the electrons inelastically and thus 
randomizes the phase of the electronic wave functions. These interactions 
try to destroy the phase coherence of the electrons which can remove the 
quantum effects. Hence, it is necessary to do the calculations at 
sufficiently low temperatures such that the phase coherence length of an 
electron exceeds the circumference $L$ of the ring.

In our calculations of low-field magnetic susceptibility at absolute zero 
temperature ($T=0$ K), we see that the current has only diamagnetic sign 
for perfect rings irrespective of the total number of electrons, while in 
the presence of impurity, it exhibits respectively the diamagnetic and 
paramagnetic sign for the rings with odd and even $N_e$. It is well known 
that at any finite temperature ($T \ne 0$ K), low-field current has a 
paramagnetic sign for the rings with even $N_e$ and in this section we 
determine that critical value of magnetic flux, $\phi_c(T)$, where 
the low-field current changes its sign from the paramagnetic to diamagnetic 
nature. In Fig.~\ref{longtemp} we display the variation of $\phi_c(T)$ as 
a function of $N_e$ (here $N_e$ is even only) for $60$-site disordered rings
($W=1$) at two different temperatures. The upper and lower curves in 
Figs.~\ref{longtemp}(a) and (b) are respectively for the rings with 
$T/T^{\star}=1.0$ and $0.5$. Figure~\ref{longtemp}(a) shows the results for 
the rings with NNH integral, while the same are plotted for the rings with 
LRH ($\alpha=1.6$) integrals in Fig.~\ref{longtemp}(b). From these results 
we can emphasize that, as the temperature increases the critical value of
magnetic flux $\phi_c(T)$, where the low-field current changes its sign from
the paramagnetic phase to the diamagnetic one, increases. 

\section{Concluding remarks}

In conclusion, we have investigated the behavior of persistent current in 
single-isolated mesoscopic rings subjected to both NNH and LRH integrals
within the tight-binding framework. Our exact numerical calculations have
shown that the current amplitude in disordered rings are comparable to that 
of ordered rings if we consider the model with LRH integrals instead of 
usual NNH integral models. This is due to the fact that higher order hopping 
integrals try to delocalize the energy eigenstates and thus prevents the 
reduction of current due to disorder in the rings. Later, we have studied 
the low-field magnetic response at $T=0$ K both for the perfect and 
disordered rings and our results have predicted that the sign of the currents 
can be mentioned precisely even in the presence of impurity in the rings. At 
the end, we have calculated the magnetic response at finite temperatures 
($T\ne 0$ K) and estimated the critical value of magnetic flux $\phi_c(T)$ 
where the low-field current changes its sign from the paramagnetic to 
the diamagnetic nature.

\end{document}